\documentclass[aps,prd,showpacs,floatfix,superscriptaddress,11pt]{revtex4-1}
\usepackage{xr}

\usepackage{amsmath,amssymb,physics}
\usepackage[toc,page]{appendix}
\usepackage{verbatim,graphicx,mathtools}
\usepackage{txfonts}
\usepackage{color}
\usepackage[all]{xy}
\usepackage{color}
\usepackage[all]{xy}
\usepackage{subfigure}
\usepackage{xcolor}
\usepackage{lipsum}
\usepackage[utf8]{inputenc}
\usepackage{scalerel}
\usepackage{mciteplus} 
\usepackage{hyperref}

\newcommand{\be}{\begin{eqnarray}}
\newcommand{\ee}{\end{eqnarray}}

\newcommand{\bmat}{\left ( \begin{array}{cc} }
	\newcommand{\emat}{\end{array} \right ) }

\def\Tr{\textrm{Tr}}

\newcommand{\half}{{1\over2}}

\newcounter{jvct}

\newcommand{\beq}{\begin{equation}}
\newcommand{\beqs}{\begin{equation*}}
\newcommand{\eeq}{\end{equation}}
\newcommand{\eeqs}{\end{equation*}}

\allowdisplaybreaks

\begin{document}

\title{Quantum Jackiw-Teitelboim gravity, Selberg trace formula, and random matrix theory}
\author{Antonio M. Garc\'\i a-Garc\'\i a}
\email{amgg@sjtu.edu.cn}
\affiliation{Shanghai Center for Complex Physics, 
	School of Physics and Astronomy, Shanghai Jiao Tong
	University, Shanghai 200240, China}

\author{Salom\' on Zacar\'\i as}
\email{szacarias@sjtu.edu.cn}
\affiliation{Shanghai Center for Complex Physics, 
	School of Physics and Astronomy, Shanghai Jiao Tong
	University, Shanghai 200240, China}

\begin{abstract}

We show that the partition function of quantum Jackiw-Teitelboim (JT) gravity, including topological fluctuations, is equivalent to the partition function of a Maass-Laplace operator of large -imaginary- weight acting on non-compact, infinite area, hyperbolic Riemann surfaces of arbitrary genus. 
The resulting spectrum of this open quantum system is semiclasically exact and given by a regularized Selberg trace formula, namely, it is expressed as a sum over the lengths of primitive periodic orbits of these hyperbolic surfaces. By using semiclassical techniques, we compute analytically the spectral form factor and the variance of the Wigner time delay in the diagonal approximation. We find agreement with the random matrix theory (RMT) prediction for open quantum chaotic systems. Our results show that full quantum ergodicity is a distinct feature of quantum JT gravity.

\end{abstract}\maketitle

\section{Introduction}
The study of quantum features of classically chaotic systems, usually termed quantum chaos, reveals a surprising 
degree of universality in the quantum dynamics. Universal features are observed specially at two time scales: the Ehrenfest time and the Heisenberg time. The former is a short time scale related to the build up of quantum corrections to the classical motion. It is characterized by the  exponential growth of certain correlation function, related to the square of commutators that measure the uncertainty of an observable. Interestingly, this exponential growth is controlled by the classical Lyapunov exponent \cite{larkin1969,berman1978}. The Heisenberg time is a long time scale, inverse of the mean level spacing, related to the time for a system to experience that the spectrum is discrete. In quantum chaotic systems, spectral correlations of neighboring eigenvalues are given by RMT \cite{wigner1951,dyson1962a,dyson1962b,dyson1962c,dyson1962d}. Physically, this means that, for sufficiently long times, the quantum dynamics is fully ergodic and only depends on the global symmetries of the system. Not surprisingly, quantum chaos ideas and techniques have been employed in many physical contexts. In nuclear physics, where they were originally introduced \cite{wigner1951,bohigas1971}, it was shown that the spectral correlations of highly excited states are well described by RMT \cite{wigner1951}. For a single particle in a random potential, and assuming that Anderson localization effects are not important, it was found analytically that the momentum uncertainty \cite{larkin1969} grows exponentially for short times and that spectral correlation are given by RMT \cite{efetov1983} for long times. For non-interacting chaotic systems with non-random deterministic motion, it was conjectured \cite{bohigas1984}, and later demonstrated \cite{sieber2001,muller2004} in the semiclassical limit, that level correlations are also given by RMT. Similarly, in the limit of zero-dimensional QCD, the spectrum of the Dirac operator is also correlated according to the RMT prediction \cite{verbaarschot1993a} for systems with chiral symmetry.

More recently, the proposal \cite{maldacena2015} of a universal bound in the Lyapunov exponent, that controls the exponential growth mentioned above and its saturation in field theories with a gravity dual, has reinvigorated the interest in quantum chaos. 
This saturation has been explicitly confirmed in the Sachdev-Ye-Kitaev (SYK) model \cite{kitaev2015,maldacena2016,jensen2016} consisting of $N$ Majoranas with infinite-range random interactions in zero spatial dimensions (see \cite{bohigas1971,french1970,mon1975,benet2003,sachdev1993,sachdev2010} for similar models with Dirac fermions). Interestingly,  the infrared description of the SYK model is controlled by the Schwarzian action, 
which is also the effective description of the boundary dynamics of JT gravity \cite{maldacena2016a,almheiri2015}. 
It has also been found  \cite{garcia2016,cotler2016} that spectral correlations of the SYK model are given by RMT even for energies close to the ground state \cite{garcia2017} where the model may have a gravity dual.

This observation of RMT spectral correlations in the SYK model together with the expected holographic duality with JT gravity, though encouraging, it is not a demonstration that the latter has similar spectral correlations. It is far from clear whether the holographic duality survives up to time scales of the order of the Heisenberg time. More importantly, the very full quantization of JT gravity, leading to a discrete spectrum so it is possible to carry out a spectral analysis, is still an open problem. Recently, a random matrix model for JT gravity has been proposed \cite{saad2019jt}, see also \cite{saad2019,arefeva2019,okuyama2019}, as a possible ultraviolet completion of the theory (see \cite{witten2019} for a generalization of this idea when fermions are included). 
More specifically, it assumes that the Hamiltonian
of the boundary theory is a random matrix. Therefore all results are obtained after an ensemble average whose physical meaning on the gravity side is not straightforward. 

One added problem for a quantization of JT gravity is that the boundary theory is Liouville quantum mechanics \cite{bagrets2016}, a one-dimensional system with a continuous spectrum which precludes any quantum chaotic behavior. Quantum chaos must therefore be necessarily a feature of the bulk, not the boundary. Indeed, it has recently been shown \cite{yang2018,kitaev2018} that bulk JT gravity maps to the dynamics of a quantum particle on the hyperbolic space in the presence of an imaginary magnetic field. This spectrum, which has been computed analytically \cite{comtet1987}, is still continuous which prevents any analysis of spectral correlations.
 
 Here we address this problem by quantizing JT gravity in a way reminiscent to Polyakov's approach to 
 string theory \cite{polyakov1987}. In JT gravity, the two-dimensional space is rigid of constant negative curvature, in the sense that it is just the solution of the classical equation of motion. The dynamics comes from the position of the boundary which acts as the physical boundary of the system. We relax the rigidness condition of the surface by allowing topological fluctuations of the geometry. This leads to a genus expansion, with genus $h \geq 1$, of non-compact Riemann surfaces of infinite area. 
We show that the JT gravity partition function can be written as a sum over genus of the partition function associated to the Maass-Laplace operator on the above surfaces in the limit of large (imaginary) weight of the holomorphic form. 
The area of the Riemann surfaces is infinite so it is possible the existence of complex resonances in the spectrum, poles of the resolvent. We found that the spectrum/resonances can be computed exactly as a sum over classical primitive periodic orbits by using the Selberg Zeta function and generalized Selberg trace formula. The resulting spectrum/resonances, which does not involve any ensemble average, only depends on classical information such as the primitive periodic orbits and the classical escape rate. We then compute, using semiclassical techniques, the spectral form factor and the variance of the Wigner time delay analytically and find that it agrees with the RMT prediction. This indicates that for sufficiently long times JT quantum gravity -in the above mentioned limit- is fully ergodic. It also suggests that quantum chaos may be one of its features at all time scales.

 
 \section{Quantization of JT gravity}
 The classical Euclidean action for JT gravity 
 is: 
\begin{equation}\label{fullaction}
I(g,\phi)=-2\pi\phi_0 \, \chi-\left[\frac{1}{2}\int _{\mathcal{M}} \sqrt{g} \phi (R+2)+\int_{\partial\mathcal{M}}\phi_b(K-1)\right],
\end{equation}
 where $\phi$ stands for the deviation of the dilaton field from its value at zero temperature, $\phi_0$, and $\chi$ is the Euler characteristic of the surface.
 The classical equation for the dilaton sets  $R = -2$ which implies $\mathcal{M}$ is a portion of rigid $AdS_2$ (Euclidean) space which we will represent using the hyperbolic half-plane model 
 $\mathbb{H}=\{z=x+iy\,\vert \, y>0\}$ with line element $ds^2=dx^2+dy^2/y^2$. With the remaining term in the action, the classical solution for the dilaton can also be easily found \cite{yang2018}. The nontrivial 
 dynamics of the model comes from the position of the physical boundary in the rigid  $\mathbb{H}$ space. Boundary conditions for the metric and dilaton along the physical boundary are given respectively by
 $ds\vert_{\textrm{bdy}}= d u { \frac{\phi_r}{\epsilon} }$ and $\phi\vert_{\textrm{bdy}}= { \frac{\phi_r}{\epsilon }}$
 where $u$ is the boundary theory time and $\epsilon$ parametrizes the distance to the $\mathbb{H}$ boundary. In the $\epsilon \to 0$ limit, where the physical boundary 
 approaches the rigid space boundary, the action in eq. (\ref{fullaction}) becomes:
 \begin{equation}
 I=-\int {\rm Sch}(\tan{\varphi(u)\over 2},u)\, du
 \end{equation}
 where ${\rm Sch}$ stands for the Schwarzian and $\varphi(u)$ expresses the bulk time as a function of the boundary time \cite{maldacena2016a}. Throughout we will refer to the $\epsilon\rightarrow 0$ limit as the {\it Schwarzian limit}.
 This action results from the spontaneous and explicit breaking of conformal symmetry, down to ${\rm SL}(2,\mathbb{R})$ symmetry, due to quantum and finite temperature effects. Different techniques and tools, from combinatorial analysis \cite{garcia2017,garcia2018a} based on the shared symmetry with the SYK model, to an explicit evaluation of the path integral \cite{stanford2017} and an mapping to a charged particle on $\mathbb{H}$ in the presence of an imaginary magnetic field \cite{yang2018,kitaev2018}, have been employed to compute spectral and thermodynamic properties of JT gravity. Here we will focus in this latter approach which can be summarized as follows (see \cite{kitaev2018,yang2018} for details). After integrating over the dilaton field and using the Gauss-Bonnet theorem, the remaining term containing the extrinsic curvature in the action of eq. (\ref{fullaction}) can be rewritten as
$\int(K-1)=\left(2\pi+A[\mathbf{x}]-L\right)$, 
 where $A[\mathbf{x}]$ is the area surrounded by the closed boundary curve of fixed length, $L=\beta\frac{ \phi_r}{\epsilon}$. 
The area term can be interpreted as the flux of a uniform {\it electric} field on $\mathbb{H}$ of strength $q=\frac{\phi_r}{\epsilon}$, or equivalently a uniform imaginary {\it magnetic} field $b=iq$. It is worth noticing that the analogy of JT gravity with the charged particle 
is strictly valid at the classical level since quantum mechanically the path integrals seemingly have different properties, as explained in detail in \cite{kitaev2018}. This ambiguity is eliminated whenever we consider the {\it Schwarzian limit}, that in the present context involves considering large $b$.
In the following, we will consider the case of finite $b$ and take the limit at the end in the expressions whenever necessary.
 
In the quantum level, the path integral quantization of JT gravity -using the analogy with the charged particle- is carried out including the constraint of trajectories of fixed proper length, which is related to the temperature.
By considering an appropriate regularization prescription, it was found \cite{kitaev2018,yang2018} that the partition function of JT gravity is equivalent to that of a charged particle on $\mathbb{H}$ in the presence of an imaginary magnetic field. The final form of the partition function is,
 \be \label{aads}
Z_{\mathbb{H}}(b,\tau) = e^{2\pi( q +\phi_0)}e^{\tau/2(1/4-b^2)}
 	\int { \cal D}{\bf x} \exp(-\int_0^\tau d\tau' \left(\half { \dot x^2 + \dot y^2 \over y^2 } -b {\dot x \over y}\right)), 
	\ee
	where the gauge $A_x=-b/y$ was used. The quantum properties of this system are then governed by the Schr\"{o}dinger equation $H\psi=E\psi$ where the Hamiltonian $2H=-D_m+b^2$ is the Maass-Laplace operator \be \label{mlo} D_m=y^{2}(\partial_x^2+\partial_y^2)-imy\partial_x \ee on $\mathbb{H}$ for automorphic forms of weight $m=2b$.  The dynamics of the particle is influenced by the non-zero magnetic field such that for sufficiently large magnetic field, in addition to the monotonous continuous spectrum, we have a discrete spectrum associated to the Landau levels on $\mathbb{H}$ \cite{comtet1987}. For an imaginary magnetic field, the spectrum of the Hamiltonian is always 
continuous \cite{comtet1987,yang2018} with energy $E=(1/4+k^2-q^2)/2$.  In the Schwarzian limit, the spectral density for the Schwarzian action is recovered \cite{kitaev2018, yang2018}. \\

The above Hamiltonian is the starting point of our analysis. We note that the quantum gravity problem defined on a rigid half hyperbolic plane $\mathbb{H}$ with a dynamical physical boundary is traded to the study of the spectral properties of 
the Maass-Laplace operator on $\mathbb{H}$,  which resulted from the solution of the classical equations of motion. Typically, a distinctive feature of quantum systems is a discrete spectrum. However, the spectrum related to the system above is absolutely continuous. We also note that the quantization of gravity cannot be complete because the geometry is still given by the solution of the classical equations of motion. In the following, we consider quantization of JT gravity that will potentially lead to a discrete spectrum of the above model and to spectral correlations given by RMT which confirm the quantum chaotic nature of JT gravity even for long times of the order of the Heisenberg time.

 A natural way to generalize the above arguments is by considering contributions of surfaces with non-trivial topology. This amounts considering topological fluctuations in the model which are not solutions of the classical equations of motion. In the functional form of the JT gravity problem eq. (\ref{fullaction}), the metric is still constrained to be of constant negative curvature at any point, although the latter enters as a delta function constraint \cite{saad2019jt}. It is then clear that the functional integral will have contributions from hyperbolic Riemann surfaces \cite{saad2019jt} with an asymptotic $AdS_2$ boundary \footnote{We will consider throughout the case of one asymptotic boundary } and arbitrary genus; $\Sigma_h,\,\, h\geq 1$.
To be more precise, contributions from bounded sub-regions of hyperbolic non-compact Riemann surfaces of infinite area and fixed boundary length where the boundary conditions of the JT gravity problem are met.
Explicitly, we integrate the dilaton along an appropriate contour to obtain

\begin{equation}
Z_{JT}=e^{2\pi\phi_0\chi}\int \mathcal{D}g_{\mu\nu}\, \delta(R+2)e^{-\phi_b\int (K-1)}.
\end{equation}
The functional integral is reduced to the boundary term, which according to the previous discussion -the genus zero case- is equivalently described by the path integral of a charged particle in an imaginary magnetic field on $\mathbb{H}$. For non-zero genus, we use Riemann uniformization to argue that the full path integral involves considering contributions of non-compact Riemann surfaces of infinite area and arbitrary genus, where a charged particle propagates. In addition, each of these contributions are weighted by a term proportional to the Euler characteristic $\chi=1-2h$. The final form of the functional integral is
 \begin{equation}\label{zft}
 \begin{split}
Z_{JT}=&Z_{\mathbb{H}}(b,\tau)+e^{\tau/2(1/4+b^2)}\sum_{h\geq1}e^{2\pi(q+\phi_0)\chi}\int_{\textrm{moduli}}\int \mathcal{D}\mathbf{x}e^{-I_h(\mathbf{x},b)},\\
=& Z_{\mathbb{H}}(b,\tau) + e^{\tau/2(1/4+b^2)}\sum_{h\geq1}e^{2\pi(q+\phi_0)\chi}\int_{\textrm{moduli}}\, Z^{h}_{_{\Gamma \setminus \mathbb{H}}}(b,\tau), 
\end{split}
\end{equation}
where $Z_{\mathbb{H}}$ is the partition function associated to the Poincare half plane eq. (\ref{aads}) and $Z^{h}_{_{\Gamma \setminus \mathbb{H}}}(b,\tau)$ the analog for the higher genus hyperbolic Riemann surfaces,  the remaining integral accounting for their moduli with metric given by the Weil-Petersson metric \cite{saad2019jt}. 
At genus zero, the quantization of this system has been studied in the literature  \cite{kitaev2018, yang2018}. 
 However the spectrum is continuous and monotonous which prevents the calculation of spectral correlations. 
By contrast, for higher genus surfaces, the resulting spectrum is amenable to a level statistics analysis. We will use spectral theory techniques concerning trace formulas in order to compute these higher genus partition functions, associated to the spectrum of the Maass Laplacian (\ref{mlo}) on the corresponding Riemann surfaces.

We will see below, using the ``equivalence'' in eq. (\ref{zft}), that the spectral density associated to $Z_{JT}$ -basically its Laplace transform- has a highly oscillating term which is expressed as a Selberg trace formula, namely, as a sum over the classical primitive periodic orbits of the surface. An explicit analytical evaluation of density-density spectral correlations, such as the  spectral form factor, will reveal that the spectrum of JT gravity is quantum chaotic and described by RMT.

\begin{figure}[h]
   	\centering
   	\resizebox{0.9\textwidth}{!}{\includegraphics{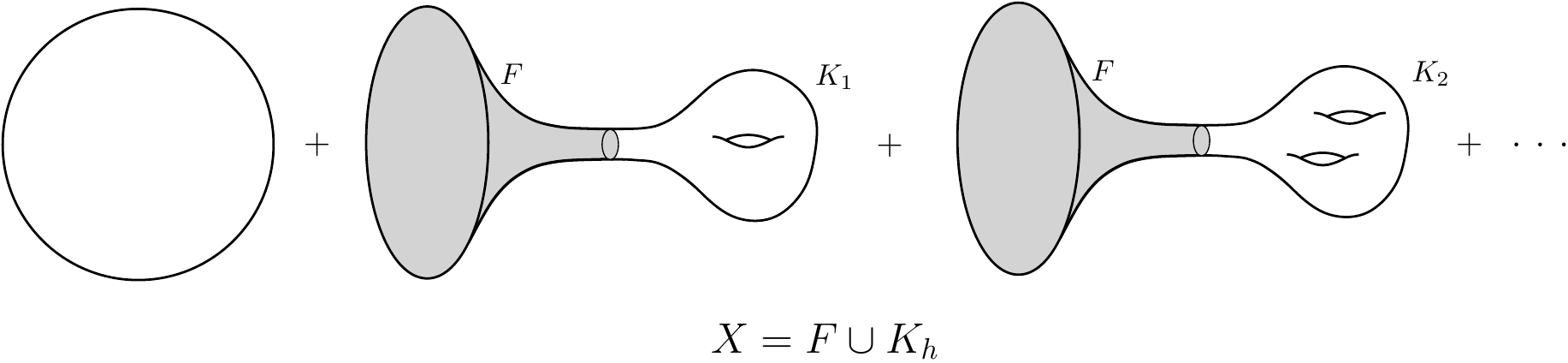}}  
   	
   	\vspace{-4mm}
   	\caption{Sketch of the leading topological corrections to the classical (left most) geometry of JT gravity. }
   	\label{fig1}
   \end{figure}

 Let us briefly clarify the notation we have used. Formally, non-compact hyperbolic Riemann surfaces of infinite area are obtained using Riemann uniformization. Namely, the Riemann surface $\Sigma_h$ is represented by a right coset of the Poincare upper half-plane model $\mathbb{H}$ by a Fuchsian group $\Gamma$ of the second kind \cite{borthwick2007}. The latter is a discrete subgroup of $PSL(2,\mathbb{R})$ with proper discontinuous action. To be more specific, we consider a fundamental domain $\mathcal{F}\subset \mathbb{H}$, the geodesic boundaries of which are paired by elements $\gamma$ of $\Gamma$ generating a surface $X=\Gamma \setminus \mathbb{H}$ homeomorphic to $\Sigma_h$. In addition to the infinite area condition, 
  the surfaces we will be considering are connected, orientable and geometrically finite (finite Euler characteristic). 
Moreover, the above Riemann surfaces are characterized by their hyperbolic ends. In the absence of cusps, geometric finiteness constraint the end space to be a funnel (infinite area elementary surface).  It is then clear that every non-compact hyperbolic Riemann surface of infinite area can be decomposed into separate pieces involving a bordered Riemann surface (compact core) and a funnel (see fig. \ref{fig1}). Explicitly, the decomposition is written as $X=F \cup K_h$ where $F$ and $K_h$ denote the funnel and compact core subregions respectively.  

Moreover, a quantity that will come very useful in the forthcoming analysis is the exponent of convergence of a Fuchsian group $\Gamma$, which is defined by
\be\label{critical}
\delta(\Gamma):= \textrm{inf}\{s\geq0: \sum_{\gamma\in \Gamma} e^{-s\, d(z,\gamma z')} <\infty\},
\ee
 where $d(z,z')$ is the distance on $\mathbb{H}$. For a Fuchsian group $\Gamma$ of the second kind and geometrically finite we have that the above exponent 
 satisfies $\delta(\Gamma)\in  (0,1)$ and equals the Hausdorff dimension of the limit set of $\Gamma$. For the infinite area surfaces, the Hausdorff dimension of the limit set 
is associated to the dynamics of geodesics which either flow around in the compact core or escape to infinity.

\subsection{Role of Moduli Space}
Notice that the equivalence between the partition function of JT gravity and $Z^{h}_{_{\Gamma \setminus \mathbb{H}}}$ -in the {\it Schwarzian limit}-
is valid up to the integral over the moduli, as seen from eq.~(\ref{zft}). For a genus $h$ Riemann surface,  there is a continuous (but finite) number of parameters characterizing (inequivalent) Riemann surfaces with the same underlying topology. The integral in eq.~(\ref{zft}) accounts for all these contributions to the JT partition function. 
As we will see below, the partition function $Z^{h}_{_{\Gamma \setminus \mathbb{H}}}$ receives contributions from continuous as well as discrete parts of the spectrum. Throughout this work we will focus on the discrete part of the spectrum. The explicit computation of the moduli integral even in this case is not straightforward since it involves performing the integration on functions depending on primitive periodic orbits of the surface \footnote{Notice that the method used in \cite{saad2019jt},  based on a decomposition of the surface as that shown in fig \ref {fig1}, accounts for the moduli using Mirzakhani's recursive method, which involves integrating functions which depend on the length of primitive {\it simple} closed geodesics instead.}. We will not attempt to address this problem here, since the aim of this work is not the detailed knowledge of the spectrum but rather the study of the dynamics of the primitive periodic orbit flow which is, in a sense, independent of Riemann surface deformations which will enable us to study the quantum chaotic features of the spectrum of  JT gravity.
More specifically, the (discrete) spectrum associated to the Riemann surfaces of genus $h$ is characterized by $\delta > 0$ defined above and the length spectrum, namely, the set of primitive periodic orbits. In all cases, it has been demonstrated that there is an exponential proliferation of long primitive geodesics \cite{naud2005,guillope1986}, which is a signature of classical chaos and also a key feature \cite{sieber2011} to establish the quantum chaotic nature of the spectrum. It is worth emphasizing that a proper understanding of other quantum features of JT gravity will necessarily involve the integration over the moduli space. 
Another feature that requires additional comment is the role of the moduli space in the computation of two-level and higher order spectral correlation functions. Here, in principle, one should have to consider the correlation of eigenvalues not only 
for a given genus $h$ surface but correlations of eigenvalues among all the set of surfaces in the moduli space. However, at least to leading order, this interference effect is likely to be negligible. The reason being that, as we shall see below, the length spectrum of each surface is unique.  According to the Selberg trace formula, the spectral correlation are expressed as products of highly oscillating sums representing different surfaces. The leading term will be given by autocorrelations, called the diagonal approximation, namely, correlations between primitive periodic orbits of a given genus $h$ surface. Even at next to leading order we expect that it will be dominated 
by correlations between different periodic orbits of the same surface. It is unclear whether correlation among geodesic of different genus $h$ surfaces could play a role for these subleading contributions but in any case it would not affect our results. 

We can summarize the discussion above by saying that, even though for other observables one must necessarily account for the contributions of all set of Riemann surfaces of given genus in the moduli space, 
this is not very relevant with respect to level statistics since all them have similar features. In this vein, we can then argue that the calculation of the path integral over geometries accepts a saddle point solution that effectively picks a Riemann surface for each topological class, defined by a certain $\delta$ and its length spectrum. Based on the above, we will be able to show that JT gravity is quantum chaotic. Our main aim is therefore to compute  $Z^{h}_{_{\Gamma \setminus \mathbb{H}}}$.
 Fortunately, this problem has been intensively investigated in both the mathematics and physics literature  \cite{selberg1956,mckean1972,bolte1993,comtet1993,dhoker1986,bolte1994,mckean1972,hejhal2006}.

\section{Evaluation of the JT partition function by the Selberg trace formula}
In this section we compute the partition function $Z^{h}_{_{\Gamma \setminus \mathbb{H}}}$  associated to the Maass Laplacian eq.~(\ref{mlo}) on the infinite area hyperbolic Riemann surfaces of genus $h$. 
Although, the spectral properties of the above surfaces are not the naive sum of the funnel and compact core contributions, we find pedagogical, and useful in order to set notation and conventions, to
briefly discuss separately the spectral features of the above subregions.

 \subsection{Spectral features of the isolated compact core and funnel} 
Let us start with the spectral features of the funnel.
In simple terms, Selberg trace formulas relate classical and quantum properties of the system. 
They contain several contributions depending on the characteristics of the Riemann surfaces and the associated spectrum of the Laplacian. However, there is a universal term given by the length spectrum $\mathcal{L}_{_{X}}$ of a given hyperbolic Riemann surface $X$ which corresponds to the set of lengths $\ell(\gamma)$ of primitive periodic orbits $\gamma$, where on the other hand the latter is an element of the conjugacy classes of $\Gamma$. Since we are just interested in the chaotic features of the system, it will be enough to discuss this contribution to the trace formula. In the case of the funnel, a regularized trace formula for non-zero magnetic field can be easily obtained following the results in \cite{borthwick2007}. This is given as the logarithmic derivative of the Selberg zeta function $Z_{_{F}}(s)$ involving the product of the length spectrum labeled by the only closed geodesic of the funnel, $\ell_{_{F}}$, together with unbounded multiplicities. This expression is not directly influenced by the non-zero magnetic field.
Even though the hyperbolic ends determine many features of the spectrum, they are not directly relevant in the study of quantum chaotic features of the system as there is only one short geodesic which is in general negligible to produce any substantial modification of short-range spectral correlations.
  
  For the compact core $Z_{_{CC}}$ -a bordered Riemann surface- the exact analytical expression of the heat kernel, equivalent to the partition function,  for non-zero magnetic field is given by the Selberg trace formula \cite{selberg1956,mckean1972,bolte1993,bolte1994,comtet1993,dhoker1986,sieber2011},
\begin{equation}
\begin{split} \label{tf}
Z_{_{CC}}\equiv\textrm{Tr}\,e^{-\tau H}=&e^{-\tau/2(1/4+b^2)}\frac{V_h}{2\pi}\int_0^\infty dk \frac{k\, e^{-k^2\tau/2}\sinh 2\pi k}{\cos 2\pi b+\cosh 2\pi k}+
\sum_{m=1}^\infty\sum_{\ell\,  \in \mathcal{L}(\gamma)} A_{\ell,m} \, g(m \ell(\gamma))e^{-\tau\,  b^2/2}, \\
&g(m \ell(\gamma))=\frac{1}{4\sqrt{\pi\tau}}e^{-\tau/8-(m \ell(\gamma))^2/4\tau},  \quad A_{\ell,m}=\frac{\ell}{2\sinh(m\ell/2)}, 
\end{split}
\end{equation}

where $H$ is the Maass Laplacian defined above, $m$ corresponds to the multiplicities of the primitive periodic orbits and $V_h=-2\pi \chi$ -front of the induced continuous contribution to the heat kernel-  is the area of the surface. This last dependence has to be removed regarding the interpretation of the partition function as that of a gravitational system, as pointed out in \cite{kitaev2018}. We notice that the explicit $b$ dependence enters as a simple shift of the case without magnetic field.
In addition, to simplify notation, we have not included explicitly the term involving the geodesic related to the border of the surface, see \cite{bolte1993} for a full expression. 
From eq.~(\ref{tf}) we see that the trace formula depends on the corresponding Fuchsian group $\Gamma$ associated to the quotient surface, which in turn depends on the parameters associated to the moduli of the hyperbolic Riemann surface. 
However, one of the strengths of the trace formula is that even though we may expect -in general grounds- a change in the length spectrum for different points in the moduli space, the partition function will still be expressed in a very compact form, eq.~(\ref{tf}). As was argued earlier, the detailed account of the moduli space is not important for our purposes since the quantum chaotic nature of the motion occurs for all Fuchsian groups associated to a given hyperbolic surface of genus $h$.
  
More importantly, unlike the Gutzwiller trace formula \cite{gutzwiller2013}, broadly used to compute the spectral density of quantum billiards in flat space, which is only valid in the semiclassical limit, the Selberg trace formula is exact, namely, the exact quantum spectrum of the system is encoded in the primitive periodic orbits of the classical counterpart. Another remarkable feature is that the dependence on the magnetic field is just a prefactor, 
of no much physical relevance. 
This feature will greatly simplify the calculation of spectral correlations associated to $Z^{h}_{_{\Gamma \setminus \mathbb{H}}}$ that we discuss now.

\subsection{Spectral features associated to  $Z^{h}_{_{\Gamma \setminus \mathbb{H}}}$}

 Having discussed the spectral properties of each separate subregion of the hyperbolic Riemann surfaces, we now study them as a whole. 
 For the case of a vanishing magnetic field, the spectrum of the total surface is relatively well understood  \cite{borthwick2007}. 

 Since the surfaces are non-compact of infinite area, it is more convenient to characterize the spectrum through resonances $s$ which are the poles of the (meromorphically continued) resolvent $R(s) = {\Tr} ((H -s(1-s))^{-1})$.
 Even in this more general situation, it is possible to show \cite{borthwick2003,patterson2001} that the zeros of the Selberg Zeta function, 
 $Z_{_{X}}(s)=\prod_{\mathcal{L}_{_{X}}} \prod_{m=0}^{\infty}(1-e^{-(s+m)\ell}),\textrm{Re}(s)>1$, provide an exact description of the quantum spectrum/resonances \cite{borthwick2005}.

 More interestingly, for the purpose of the calculation of spectral correlations, the partition function, and therefore the spectral density, take the form of a regularized Selberg trace formula \cite{borthwick2011} 
 which is given by the length spectrum $\mathcal{L}_{_{X}}$ of the surface where the contribution of each primitive periodic orbit is still given by $A_{\ell,m}$,  as in the compact case eq.~(\ref{al}). However, we note that 
 this length spectrum is in principle completely different from the one corresponding 
 to the compact core. For instance, we expect that in this case some geodesics originally in the compact core are now missing as the surface is now of infinite area and therefore corresponding to that of an open system, where it is possible for them to escape outside the compact core.
 Regarding the dependence on the magnetic field, we do not expect a qualitative dependence because the Selberg trace formula above is largely independent of it though no firm conclusion can be achieved until a full analysis of this dependence is carried out.

 Even though the spectrum of the Maass-Laplace operator can in principle be computed from information of the length spectrum $\mathcal{L}_{_{X}}$, explicit calculations of the latter for infinite area surfaces are scarce, see \cite{borthwick2014} for a surface with $h = 1$. 
 However, our main aim is not a detailed knowledge of the spectrum but rather to clarify whether the Maass-Laplace spectrum, or correspondingly the JT spectrum, is quantum chaotic with spectral correlations described by RMT.

 An important quantity which helps answering this question is the the Hausdorff  dimension $\delta$ of the classical attractor (trapped set) \cite{borthwick2007,beardon1968,sullivan1984,patterson1976} introduced in eq. (\ref{critical}). This dimension is directly related to the escape rate $\upsilon = 1- \delta$, the rate at which trajectories close to the trapped set escape to infinity in units of the inverse of the Heisenberg time $\tau_{_{H}}$.

 Crucial for the forthcoming spectral analysis, is the fact that for all hyperbolic Riemann surfaces of infinite area, $\delta > 0$ \cite{beardon1968,borthwick2007}.  Its specific value will be highly dependent on other parameters defining the surface, such as the number of genus, but it is always positive.

 The finiteness of $\delta$ is directly related to other relevant results:
 
 \begin{itemize}
 	\item For $\delta > 0$, the number of closed geodesics of length less than $t$, for $t$ sufficiently large, grows exponentially $\sim e^{\delta t}/\delta t$ \, \cite{naud2005,guillope1986}. This is a generalized version for hyperbolic Riemann surfaces of infinity area of the so called {\it prime geodesic theorem}.
 	
 	\item  The eigenvalues with the largest real part is $\delta$ and there are no other resonances in the line ${\rm Re} s = \delta$ \cite{patterson1976,patterson1988}. 
 	
 	

 	\item  The topological entropy $S_T$ of the geodesic flow of trapped set is positive $S_T = \delta$ \cite{patterson2006,keane1991}. 
 \end{itemize}
 
 The finite topological entropy is a distinctive feature of classical chaos while 
 the exponential growth of long periodic orbits play a key role for the 
 demonstration of quantum chaotic features in the spectrum.  
 
 In summary, taking into account that the partition function is still given by just the length spectrum $\mathcal{L}_{_{X}}$ \cite{borthwick2005}, with amplitudes $A_{\ell,m}$ similar to the compact case eq.~(\ref{al}), and that the classical dynamics is chaotic, we broadly expect that quantum JT gravity is also quantum chaotic, namely, its spectral correlations are well described by RMT. In next section we provide evidence that this is the case.

   \section{Spectral correlations of JT gravity: Spectral form factor and Wigner time delay fluctuations}
Having  investigated the features of the partition function of JT gravity, we now move to the calculation of spectral correlations in order to confirm agreement with the predictions of RMT. For that purpose, the first step is to relate the JT partition function with the spectral density $\rho(E)=\sum_{i}\delta(E-E_i)={\bar \rho} + {\tilde \rho}(E)$ 
where the first term stands for the monotonous part whilst the second stands for the oscillating part. We note that the spectrum is in general complex so the spectral density is smoothed out with respect to a Dirac delta function. In general, only the latter enters in observables relevant to establish the quantum chaotic nature of the spectrum. Fortunately, the spectral density is nothing but the Laplace transform of the partition function $\rho(E)\sim \int _{_{C}} d\tau\,  Z(\tau)\, e^{E\tau}$ with $C$ a vertical path in the complex plane.  Taking into account that the JT partition function -given by the trace formula- is also naturally split into a monotonous and 
oscillating part, it is straightforward to show that \be \label{fsd} {\tilde \rho}(E) \sim \sum_{\mathcal{L}_{_{X}}} A_\ell e^{ik\ell}, \ee
where $E \sim k^2$, and \be \label{al} A_\ell \sim \frac{\ell}{2\sinh(\ell/2)},\ee 
where for convenience we have not included multiplicities $m$ as they lead to subleading corrections in the correlations. We note that different length spectra leads to different attractor dimensions $\delta > 0$. We recall also that the length spectrum
for the whole surface is not in principle related to that of the compact core eq.~(\ref{tf}).

In order to characterize the nature of the quantum dynamics we will investigate 
the spectral form factor and the Wigner time delay fluctuations. 
The latter is an observable employed to characterize the spectrum of open quantum chaotic systems. It is relevant in our case since $0 <\delta < 1$ and therefore the escape rate $\upsilon > 0$.


\subsection{Spectral form factor}
The simple form corresponding to the oscillating part of the spectral density -given in terms of a sum over primitive periodic orbits- makes possible the analytical calculation of spectral correlations by using semiclassical techniques. Our aim is to show explicitly that level statistics agree with the RMT prediction which is a signature of quantum chaos. 
If the escape rate $\upsilon$ is negligible, namely, many closed geodesics stay in the compact core, then, the lowest part of the spectrum is discrete and the system is effectively closed. In that case, the spectral form factor $K(\tau)$, the Fourier transform of the two-level correlation function, is a good indicator of quantum chaos. Despite the (imaginary) magnetic field, which breaks time reversal invariance, our model fall within the GOE universality class typical of time reversal systems because the sum over primitive periodic orbits in eq.~(\ref{tf}) is doubly degenerate which mimics the effect of time reversal invariance  \cite{comtet1993}. In this case, the GOE result for closed systems is $K(\tau) = 2\tau$ for $\tau \ll 1$. We show next that the spectral form factor for JT gravity agrees with the RMT prediction in this limit and derive the extension of this result in the case of a finite escape ratio $\upsilon$.

The spectral form factor is explicitly given by

\be
K(\tau)=\left  \langle \int_{-\infty}^\infty \frac{d\eta}{{\bar \rho}(E)} \langle {\tilde \rho}(E+\eta/2){\tilde \rho}(E-\eta/2)  e^{i2\pi \tau \eta{\bar \rho}(E)}\right \rangle_E,
\ee

where ${\bar \rho}(E)$ is the Heisenberg time in our system, the inverse of the mean level spacing. Inserting the analytical expression for the density eqs.~(\ref{fsd}),~(\ref{al}), results in \cite{sieber2001},
\begin{equation}
K(\tau)=\frac{1}{T_{_{H}}} \left\langle\sum_{\ell,\ell'\in \mathcal{L}_{_{X}}}A_\ell A_{\ell^{'}} 
e^{i(S_\ell - S_\ell')/\hbar} \delta \left(T-\frac{\ell+\ell'}{2}\right) \right\rangle_{E}
\end{equation}
where,  $S_\ell \sim k\ell$ is the classical action related to the geodesic $\ell$ and momentum $k$.    
Off-diagonal terms $\ell \neq \ell^{'}$ are suppressed due to both the sum over quasi random phases and the averaging procedure so we only consider primitive periodic orbits without repetition.  With these simplifications, the spectral form factor is given by, 
\begin{equation}
K(\tau) = \frac{2}{T_H}\sum_{\ell\in \mathcal{L}_{_{X}}} A_\ell^2\delta(T-\ell) 
\end{equation}
In the semiclassical limit of interest, the integral is dominated by long periodic orbits, so, using eq.~(\ref{al}), $A_\ell \approx \ell/e^{\ell/2}$. The sum is then replaced by an integral where the density of periodic orbits is the derivative of the number of periodic orbits of length less than $L'$ 
which according to the {\it prime geodesic theorem} \cite{naud2005,guillope1986} is $\sim e^{\delta L'}/L'$. Therefore the integral simplifies to:
\begin{equation}
K(\tau) \approx \frac{2}{\tau_H}\int_{-\infty}^{\infty}dL' L' e^{-(1-\delta) L'}\delta(T-L')\sim \frac{2T}{T_H}e^{-(1-\delta)T} = 2\tau e^{-\upsilon \tau},
\end{equation}
where $T/T_H = \tau$, $\upsilon = (1-\delta)T_H$ is the escape rate in units of the mean level spacing, which agrees with the random matrix result in the limit $\delta \to 1$ corresponding to a closed quantum chaotic system. For $\delta \neq 1$, it also agrees with the prediction for an open quantum chaotic system in the semiclassical limit \cite{vallejos1998,eckhardt1993,kuipers2007} or a random matrix scattering matrix \cite{savin1995}. We note that since $\delta$ is the lowest eigenvalue/resonance, $T_H \sim f(\delta)$ with $T_H \to \infty$ for $\delta \to 0$ close to the ground state.

The calculation of higher order terms in the $\tau$ expansion is feasible though rather cumbersome. In the limit $\delta \to 1$, it has been carried out in Ref.~\cite{sieber2011} and it agrees with the RMT prediction. We expect that for $\delta \neq 1$, it would also agree with previous results from RMT and semiclassical open quantum chaotic systems  \cite{kuipers2007}. 

 Note that $\tau$ is measured in units of the Heisenberg time so even though it is an expansion in small $\tau$, it describes the time evolution of the system for long times, of the order but smaller than the Heisenberg time.

\subsection{Wigner time delay fluctuations}
Since our system is open and we do not know the precise 
value of $\delta$ in our case, it is in principle not clear to what extent it is possible to demonstrate the existence of quantum chaos from the spectral fluctuations for sufficiently small $\delta$ where it may not be possible to distinguish single eigenvalues. We note however that there are observables that signal quantum chaotic features in open chaotic systems. One of the most popular is the Wigner time delay $\tau_W$ \cite{wigner1955,smith1960} defined as the extra time a scattering process takes with respect to free motion. Alternatively, it can also be defined as the difference between the spectral density of the open scattering system, that includes resonances (poles of the resolvent), and a free system. Using the expression for the spectral density eq.~(\ref{fsd}), it is given by,
\be
\label{wdtpo}
\tau_W \approx {\bar \tau}_W + \frac{2}{\upsilon \tau_{_{H}}}{\rm Re}\sum_{m=0}\sum_{\ell\in \mathcal{L}_{_{X}}}A_{\ell,m} e^{\frac{i}{\hbar}mS_\ell}
\ee
where ${\bar \tau_W}$ is the smooth part of the time delay that does not enter in the calculation of fluctuations, $A_{\ell,m}$ was defined in eq.~(\ref{al}) and $S_\ell \propto \ell$ is the classical action related to the primitive geodesic of length $\ell$ and $\upsilon$ is the escape rate defined above.
A RMT prediction, based on the modeling of the scattering matrix as a random matrix,  for the variance and, among others, energy fluctuations of $\tau_W$ is available \cite{lehmann1995,fyodorov1997}. Assuming that the time scale related to the shortest periodic orbit is the smallest length scale in the problem, it agrees with that of deterministic quantum chaotic systems by using the trace formula \cite{kuipers2007,kuipers2008,brouwer2007,eckhardt1993}. Interestingly, the same results are obtained by using only the periodic orbits inside the scattering region, or including the full orbits that eventually escape from it, see Ref.~\cite{kuipers2014} and references therein.

A useful indicator of quantum chaos is the Wigner time delay variance ${\rm var}\,  \tau_W$, given by \cite{kuipers2014,kuipers2007,kuipers2008}, 
\be
{\rm var}(\tau_W) \approx \frac{2}{\tau_{_{H}}^2}{\rm Re} \left \langle \sum_{\ell,\ell'\in \mathcal{L}_{_{X}}}A_\ell A_{\ell'}^{*}e^{i(S_\ell -S_{\ell'})/\hbar}   \right \rangle ,
\ee
where only $m = 1$ terms, this corresponds to neglecting repetitions, are considered because higher $m$ contributions are exponentially smaller. 

As in the calculation of the spectral form factor, the leading term in the semiclassical approximation corresponds to the diagonal approximation $\ell = \ell'$. The resulting single sum can be efficiently evaluated by using the prime geodesic theorem  \cite{naud2005,guillope1986}, and the expression for $A_\ell$ eq.~(\ref{al})
in the limit of large $\ell$. That results in,
\be
{\rm var} (\tau_W) = \frac{2}{\tau_H^2\upsilon^2}+\ldots ,
\ee
with $\tau_H$ the Heisenberg time, the time scale related to the mean level spacing. This in agreement with the RMT prediction. Indeed agreement has been found up to $8$th order in the expansion parameter $1/\tau_H\upsilon$ \cite{kuipers2007}. We note that this expansion parameter is a sensible choice because if $\upsilon \tau_H \geq 1$ the spectrum could at all effects be considered discrete, as in a closed system, where observables like the spectral form factor are more suitable to describe the quantum motion.

Agreement with RMT predictions is also found for other observables such as energy correlation of the time delay or other correlations involving the scattering matrix \cite{kuipers2014,savin2014}.


The main goal of the paper was to show that the result of a quantization of JT gravity whenever topological fluctuations are allowed,  is that the dynamics is quantum chaotic. Namely, spectral correlations are given by RMT for long time scales of the order of the Heisenberg time. 
The results of this section strongly suggest that this is the case.
 We stress that key feature for this finding is that for any $\delta > 0$ there is an asymptotic exponential growth \cite{naud2005,guillope1986} of primitive periodic orbits leading to a finite topological entropy. This property guaranties the analytical calculation of level statistics by using semiclassical techniques. We therefore expect that the obtained random matrix correlations are a robust feature of the spectrum of the Maass Laplacian on any of the hyperbolic Riemann surfaces of infinite area in the moduli space of the theory because in all of them $\delta > 0$.
  

\section{Discussion and Conclusion}

In this exploratory study, we have studied the quantization of JT gravity whenever topological fluctuations are allowed. 
We have shown that the original quantum gravity problem is mapped onto the calculation of the spectrum of a certain Maass-Laplace differential operator on non-compact Riemann surfaces of infinite area and genus $h \geq 1$ Remarkably, the spectrum of this open chaotic system is semiclassically exact. The spectrum and resonances of this operator are written explicitly by a Selberg trace formula, namely, a sum over geodesics of the classical counterpart. Resonances, corresponding to classical trajectories escaping to infinity, are zeros of the associated Selberg theta function. The spectral form factor, the variance of the Wigner time delay agrees with the RMT prediction. This is an indication that full quantum ergodicity is a distinctive feature of quantum JT gravity. We stress that a requirement to observe quantum chaos is that the time scale related to shortest closed geodesic, a sort of Thouless time, must be much shorter than the inverse of the classical escape rate $\gamma = 1 - \delta$ with $\delta >0$. These quantities will depend among others on the dilaton boundary value, which is proportional to the {\it magnetic} field, and the genus $h \geq 1$ of the surface.
It would be interesting to carry out an explicit calculation of $\delta$ and $\gamma$ to fully confirm the quantum chaotic nature of quantum JT gravity. Moreover, the analogy between quantum JT gravity and the charged particle picture is only fully justified \cite{yang2018,kitaev2018} in the Schwarzian limit corresponding to a large and imaginary magnetic field. Therefore this is another condition for our results to hold.

Other topics that deserve further attention are the generalization to higher spatial dimensions, the calculation of non-universal corrections to random matrix results for sufficiently short times and, following the ideas of \cite{witten2019}, the generalization of the results presented here to the supersymmetric case,  which we plan to report in the near future.

\acknowledgments
A.M.G.G thanks Dario Rosa, Jac Verbaarschot and Zhenbin Yang for illuminating discussions. We acknowledge financial support from a Shanghai talent program and from the National Natural Science Foundation of China (NSFC) (Grant number 11874259)

\bibliography{library2}
 
\end{document}